\documentclass[useAMS,usenatbib,usegraphicx]{mn2e}

\title[MegaZ-LRG]{MegaZ-LRG: a photometric redshift catalogue of one million SDSS Luminous Red Galaxies}
\author[A. Collister, O. Lahav, C. Blake et al.]{Adrian Collister$^{1}$, Ofer Lahav$^{2}$\thanks{lahav@star.ucl.ac.uk}, Chris Blake$^{3,4}$, Russell Cannon$^{5}$, \newauthor Scott Croom$^{5}$, Michael Drinkwater$^{6}$, Alastair Edge$^{7}$, Daniel Eisenstein$^{8}$, \newauthor Jon Loveday$^{9}$, Robert Nichol$^{10}$, Kevin Pimbblet$^{6}$, Roberto De Propris$^{11}$, \newauthor Isaac Roseboom$^{6}$, Nic Ross$^{7}$ , Donald P. Schneider$^{12}$, Tom Shanks$^{7}$, David Wake$^{7}$
\\$^{1}$ Institute of Astronomy, University of Cambridge, Cambridge CB3 0HA, UK 
\\$^{2}$ Department of Physics \& Astronomy, University College London, Gower Street, London WC1E 6BT, UK 
\\$^{3}$ Department of Physics \& Astronomy, University of British Columbia, 6224 Agricultural Road, Vancouver, B.C., V6T 1Z1, Canada
\\$^{4}$ Centre for Astrophysics \& Supercomputing, Swinburne University of Technology, P.O.Box 218, Hawthorn, VIC 3122, Australia
\\$^{5}$ Anglo-Australian Observatory, PO Box 296, Epping, NSW 1710, Australia 
\\$^{6}$ Department of Physics, University of Queensland, Brisbane, QLD 4072, Australia 
\\$^{7}$ Department of Physics, University of Durham, South Road, Durham DH1 3LE, UK 
\\$^{8}$ Steward Observatory, University of Arizona, 933 N. Cherry Ave, Tucson, AZ 85721, USA 
\\$^{9}$ Astronomy Centre, University of Sussex, Falmer, Brighton BN1 9QJ, UK 
\\$^{10}$ Institute of Cosmology and Gravitation, University of Portsmouth, Portsmouth PO1 2EG, UK 
\\$^{11}$ Cerro Tololo Inter-American Observatory, Casilla 603, La Serena, Chile 
\\$^{12}$ Department of Astronomy \& Astrophysics, The Pennsylvania State University, 525 Davey Laboratory, University Park, PA 16802, USA}

\date{}

\pagerange{\pageref{firstpage}--\pageref{lastpage}} \pubyear{2006}

\begin{document}

\label{firstpage}

\maketitle

\begin{abstract}
We describe the construction of MegaZ-LRG, a photometric redshift catalogue of over one million luminous red galaxies (LRGs) in the redshift range $0.4<z<0.7$ with limiting magnitude $i<20$. The catalogue is selected from the imaging data of the Sloan Digital Sky Survey Data Release 4. The 2dF-SDSS LRG and Quasar (2SLAQ) spectroscopic redshift catalogue of  13,000 intermediate-redshift LRGs provides a photometric redshift training set, allowing use of ANNz, a neural network-based photometric-redshift estimator. The rms photometric redshift accuracy obtained for an evaluation set selected from the 2SLAQ sample is $\sigma_z = 0.049$ averaged over all galaxies, and $\sigma_z = 0.040$ for a brighter subsample ($i < 19.0$). The catalogue is expected to contain $\sim$5 per cent stellar contamination. The ANNz code is used to compute a refined star/galaxy probability based on a range of photometric parameters; this allows the contamination fraction to be reduced to 2 per cent with negligible loss of genuine galaxies. The MegaZ-LRG catalogue is publicly available on the World Wide Web from \texttt{http://www.2slaq.info}. 
\end{abstract}

\section{Introduction}

Galaxy redshift surveys have been a cornerstone amongst probes of the Universe since Hubble's discovery of the cosmological expansion in 1929. Recent years have witnessed the construction of exquisitely detailed maps of the local ($z\sim0.1$) Universe by the 2-degree Field Galaxy Redshift Survey (2dFGRS; \citealt{colless}) and the Sloan Digital Sky Survey (SDSS; \citealt{york}). These surveys have defined the new state-of-the-art in spectroscopic detector technology, each constructing spectrographs capable of simultaneous observation of hundreds of objects. However, further significant increases in the depth and area accessible to spectroscopic redshift surveys will rely on the development of a new generation of instruments. 

Photometric redshifts, which are estimated from broadband galaxy colours rather than spectra, offer an invaluable interim solution. Relative to multi-object spectroscopy, high-quality photometry can be obtained far more quickly and for significantly fainter sources. Photometric redshift estimators are numerous, but generally involve calibration against either an observed spectroscopic training set (e.g. polynomial fitting, \citealt{connolly}, or neural networks, \citealt*{firth}), or a small set of template spectra (e.g. HYPERz; \citealt*{bolz}; see also \citealt{benitez}, \citealt{csabai}). The accuracy of photometric redshifts will never approach the precision possible with spectroscopic redshifts, but the efficiency of the method allows vastly wider and deeper surveys to be conceived. 

This paper describes the construction of the MegaZ-LRG photometric redshift catalogue. MegaZ-LRG comprises more than one million intermediate-redshift ($0.4\la z\la 0.7$) luminous red galaxies (LRGs) selected from the imaging data of the SDSS Data Release 4 (DR4; \citealt{adel}). Lower redshift LRGs ($z\la0.45$) are already targetted with the SDSS spectrograph across the SDSS survey area \citep{eis01}. By making use of the photometric redshift technique, MegaZ-LRG provides redshift information (albeit less accurate than that provided by spectroscopic study) for significantly more distant and for a far greater number of LRGs. LRGs are particularly suited to the photometric redshift technique due to the homogeneity of the population \citep{eis03} and, especially, the prominence of the 4000\AA~ break in their spectra. 

\begin{figure}
\includegraphics[width=0.47\textwidth]{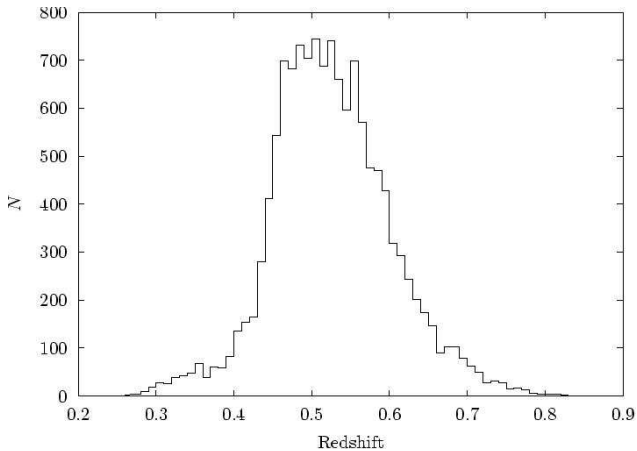}
\caption{Spectroscopic redshift distribution of the 2SLAQ LRG sample (13,115 galaxies), in redshift bins of width $\Delta z = 0.01$.}
\label{fig1}
\end{figure}

The construction of the MegaZ-LRG catalogue has been facilitated by the recent completion of the 2dF-SDSS LRG and Quasar survey (2SLAQ; \citealt{cannon}). 2SLAQ combined the high-precision SDSS imaging with the exceptional spectroscopic capabilities of the Two-degree Field (2dF) instrument on the 3.9-m Anglo-Australian Telescope to produce a spectroscopic redshift catalogue of $\sim$13,000 luminous red galaxies in the redshift range $0.4<z<0.7$ (Figure \ref{fig1}).

\begin{figure*}
\includegraphics[angle=270,width=\textwidth]{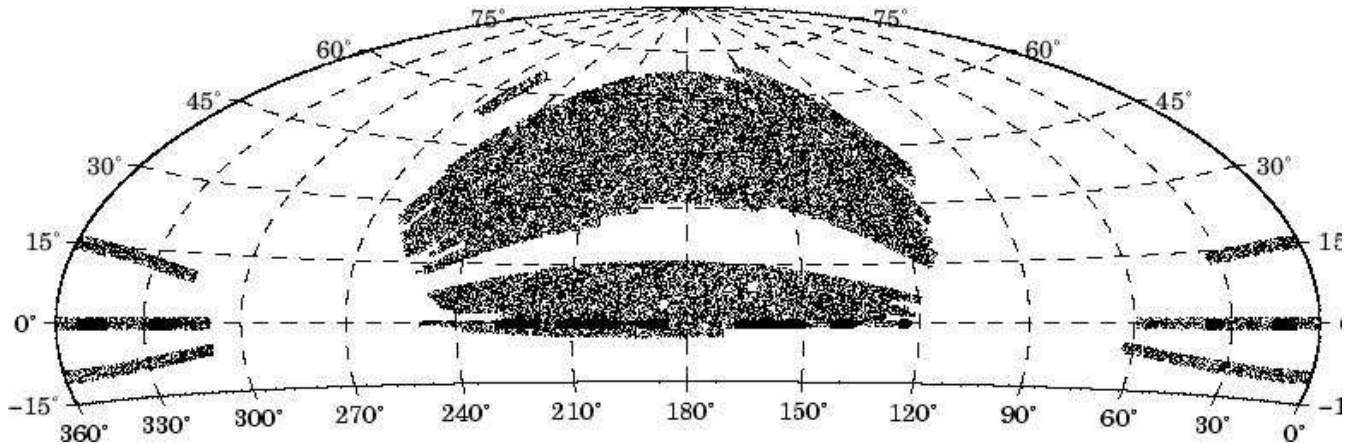}
\caption{\label{fig2}Map of MegaZ-LRG sample (small points covering the entire SDSS DR4 area), and 2SLAQ fields (black regions centred on equator). For clarity, only 50,000 randomly-selected MegaZ-LRG galaxies are shown. Equal-area Aitoff projection of equatorial coordinates.}
\end{figure*}

By necessity, the 2SLAQ survey was restricted to a limited number of fields located in the equatorial stripe of the SDSS survey area (Figure \ref{fig2}). Applying the 2SLAQ LRG selection to the entire SDSS DR4 imaging area returns a sample of over one million galaxies. The spectroscopic 2SLAQ catalogue constitutes a superb photometric redshift training set for this sample. The value of such a training set is twofold: most importantly, it enables a detailed analysis of the photometric redshift error distribution, but it can also be used to calibrate the photometric redshift estimator. We make use of ANNz \citep{collister}, a neural network-based photometric redshift estimator, for which the existence of a well-representative training set is essential.

The structure of this paper is as follows. In the next Section we describe the criteria used to select the MegaZ-LRG target sample from the SDSS DR4 imaging catalogue. In Section \ref{sec.photoz} we explain the ANNz photometric redshift technique and evaluate the accuracy of the photometric redshifts obtained for the MegaZ-LRG catalogue. In Section \ref{sec.sg} ANNz is used to refine the star/galaxy separation in the catalogue. Finally, Section \ref{sec.megazlrg} describes the MegaZ-LRG catalogue itself.

\section{Target selection}
\label{sec.selection}

MegaZ-LRG is based on SDSS five-band ($ugriz$; \citealt{fuku}, \citealt{smith}) imaging data obtained with a large format CCD camera \citep{gunn98} mounted on a special-purpose 2.5-m telescope \citep{gunn06} located at Apache Point Observatory in New Mexico. The photometric accuracy is on the order of a few percent, and the astrometric accuracy of the object positions is approximately 0.1\arcsec\ \citep{pier}. Technical details can be found in \citet{york} and \citet{stoughton}.

\subsection{Selection criteria}
\label{sec.criteria}

The neural-network technique for photometric redshift estimation relies on the training set being well-representative of the target sample. Our catalogue selection is therefore based directly on that of the 2SLAQ LRG sample. The 2SLAQ selection criteria for identifying LRGs at $0.4 < z < 0.7$ are described in the following subsections (see \citealt{cannon} for a more detailed explanation of the selection). All magnitudes are corrected for Galactic extinction following \citealt*{schlegel}. The various magnitude types are described in Table \ref{tab.mags}.

\begin{table*}
\centering
\caption{Definitions of the different types of SDSS magnitude used in the MegaZ-LRG selection.}
\label{tab.mags}
\begin{tabular}{ll}
\hline
Type of magnitude & Definition\\
\hline
psf magnitude & Magnitude corresponding to the best fit of the point-spread function \\
& at the galaxy position. Useful for star-galaxy separation.\\
de Vaucouleurs magnitude & Magnitude corresponding to the best fit of a de Vaucouleurs profile.\\
& The best estimate of the total flux for faint LRGs.\\
exponential magnitude & Magnitude corresponding to the best fit of an exponential profile.\\
model magnitude & Uses the best fit of a de Vaucouleurs or exponential profile in the r-band,\\
& with the amplitude scaled to fit measurements in other filters. \\
& This is the best estimator of the colour of the galaxy, because the same \\
& aperture is used for all the filters.\\ 
fibre magnitude & The flux contained within the aperture of a spectroscopic fibre (3\arcsec in diameter).\\
\hline
\end{tabular}
\centering
\end{table*}

Magnitude limits are motivated primarily by the need for sufficient flux for the 2dF spectrograph. The de Vaucouleurs magnitude provides the best measure of the total flux for faint LRGs.

\begin{equation}
i_\rmn{fibre} < 21.4; 
\end{equation}
\begin{equation}
17.5 < i_\rmn{deV} < 20.0.
\end{equation}

The 2SLAQ catalogue has high ($\sim$90 per cent) completeness to $i_\rmn{deV} < 19.8$, but drops off sharply beyond this limit (Figure \ref{fig3}). In fact the nominal 2SLAQ flux limit is $i_\rmn{deV} = 19.8$; the small number of objects fainter than this were obtained in an early observing run in which the flux limit was temporarily moved to $i_\rmn{deV} = 20.0$ (see \citealt{cannon} for details). 

Colour cuts are used to isolate the LRGs. All colours are calculated using model magnitudes; these provide unbiased colours since they are based on an identical aperture in every band. The colour selection is illustrated by Figure \ref{fig4}.
\begin{equation}
0.5 < g - r < 3; 
\end{equation}
\begin{equation}
r - i < 2;
\end{equation}
\begin{equation}
c_\rmn{par} \equiv 0.7(g - r) + 1.2(r - i - 0.18) > 1.6;
\end{equation} 
\begin{equation}
d_\rmn{perp} \equiv (r - i) - (g - r)/8.0 > 0.5.
\end{equation}
The selection on $c_\rmn{par}$ separates later-type galaxies from the LRG sample. Cuts above lines of constant $d_\rmn{perp}$ select early-type galaxies with increasingly high redshift. The main 2SLAQ sample is defined by $d_\rmn{perp} > 0.55$, but a small number of (lower-redshift) LRGs were observed below this boundary during the initial observing runs.

Effective star-galaxy separation is performed using the following criteria (see also Section \ref{sec.sg}):
\begin{equation}
i_\rmn{psf} - i_\rmn{model} > 0.2~(21.0 - i_\rmn{deV} ); 
\end{equation}
\begin{equation}
$i-band de Vaucouleurs radius$ > 0.2\arcsec. 
\end{equation}
Note that the SDSS star-galaxy classification is not used.

Some final technical requirements: ``Detected'' in both $r$ and $i$; \texttt{nchild} = 0; not \texttt{SATURATED} in any band; not \texttt{NO\_PETRO} in $r$ or $i$. 

These selection criteria are extremely effective at identifying LRGs in the redshift range of interest: 95 per cent of the objects targeted by 2SLAQ are \emph{bona fide} intermediate-redshift LRGs. The most significant contaminant, accounting for virtually all of the remaining 5 per cent, are M-type stars. These cannot be trivially separated from the LRGs using $gri$ colours, and the small angular diameters subtended by galaxies at these distances mean that it is difficult to distinguish between LRGs and point-spread functions morphologically. In Section \ref{sec.sg} we make use of additional photometric parameters to derive an enhanced, neural network-based star/galaxy separation flag.

\subsection{The photometric target sample}

\begin{figure}
\includegraphics[width=0.47\textwidth]{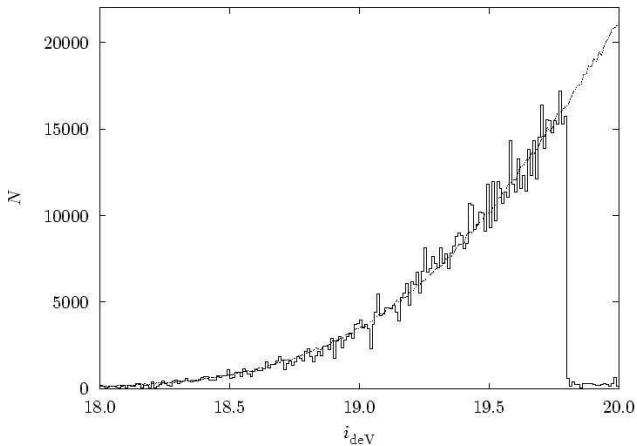}
\caption{Distribution of $i$-band de Vaucouleurs magnitude for the MegaZ-LRG target sample (dotted line) and the 2SLAQ training set (histogram). The 2SLAQ histogram is scaled so that the two distributions have equal integrated area in the region $i_\rmn{deV} < 19.8$. The bin size is $i_\rmn{deV} = 0.01$.}
\label{fig3}
\end{figure}

\begin{figure}
\includegraphics[width=0.47\textwidth]{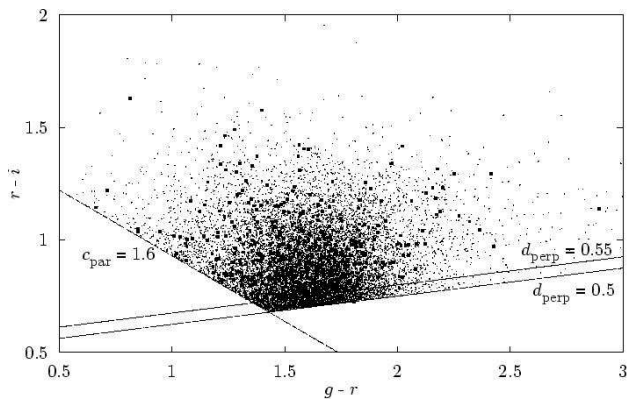}
\caption{Colour distributions of 2SLAQ LRGs (squares) and MegaZ-LRG targets (small points). For clarity, only a randomly-selected subset of each catalogue is shown.}
\label{fig4}
\end{figure}

The MegaZ-LRG photometric sample is selected from the SDSS DR4 imaging catalogue using the criteria exactly as specified above. Only \texttt{PRIMARY} objects are included in order to omit duplicate observations. Submitting the selection criteria to the SDSS DR4 catalogue returns 1,214,117 objects. The de Vaucouleurs magnitude and model colour distributions of the sample are compared with those of the 2SLAQ LRG catalogue in Figures \ref{fig3} and \ref{fig4}.

\subsection{The training set}

The final 2SLAQ LRG catalogue provides reliable spectroscopic redshifts for 13,768 unique objects. In order to ensure consistency between the training and target samples, we obtained the DR4 photometry for the 2SLAQ objects by coordinate-matching against our photometric target sample. Confident matches were found for 13,139 objects; changes introduced into the SDSS photometric pipeline since the 2SLAQ targets were selected mean that the remainder now fail the selection.

\section{Photometric redshifts} 
\label{sec.photoz}

\subsection{ANNz}

The ANNz photometric redshift code\footnote{The ANNz software package may be obtained from \texttt{http://www.star.ucl.ac.uk/$\sim$lahav/annz.html}} (\citealt{collister}; \citealt{phd}) is based on neural networks. In common with other ``empirical'' photometric redshift estimators, it relies on the existence of a training set of objects with spectroscopic redshifts. This sample should be representative of the target photometric sample in terms of magnitude and colour-space distributions. Given a well-matched training set, the neural-network method is highly competitive with commonly-used photometric redshift estimators: it is less prone to systematic errors than the SED-fitting approach, and is found to provide the greatest accuracy amongst similar training-set-based methods (\citealt{firth}; \citealt{csabai}). The primary drawback of the method is the need for observational training data, which can be expensive to obtain. One can only apply the trained network to target objects which lie within the parameter space sampled by the training data. The technique is, therefore, not well-suited to the traditional use of photometric redshifts at faint magnitudes, where obtaining a sufficiently large training set is likely to be impossible. Rather, its strength is in producing very large redshift samples from the combination of a modest spectroscopic survey and a much wider photometric sample.

The artificial neural network (ANN) is in essence a highly-flexible, fully non-linear fitting function. The inputs to the function are the photometric parameters (usually the galaxy magnitudes in each of a range of filters), and the output is the redshift. The ANN function incorporates a number of free parameters known as weights; these are optimised (the network is ``trained'') using the training set. The training process involves minimising a ``cost function'': essentially the sum over the training set of the squared differences between the photometric and spectroscopic redshifts. The number of free parameters is controlled by the network architecture. \citet{firth} investigate the influence of network architecture on performance. For the same number of parameters, adding extra hidden layers is found to give greater gains than widening existing layers. As the network complexity is increased, the accuracy eventually converges so that no further improvement is gained by adding additional nodes. The network architectures used in the following applications are chosen (by trial-and-error) to be sufficiently complex for such convergence to be achieved. Note that the cost function includes a ``weight decay'' term that prevents weights becoming large unless they contribute a significant improvement to the performance of the network.

In practice, the available training data is subdivided into ``training'' and ``validation'' samples. Only the training set is directly used to train the network, but at each iteration of the minimiser the cost function is also evaluated on the validation set. This prevents over-fitting to the training data, by halting the training process once convergence is observed for the validation set. Once training is complete, the target galaxies are submitted to the network in turn, and the output from the network in each case is assigned as the photometric redshift. The accuracy is improved by independently training a number of networks (on the same training data) and using the mean of their outputs as the photometric redshift for each target galaxy. However, for a well-matched training set (as we have here) the gain in accuracy when using such a ``committee'' is usually minor. Full details of the ANNz software may be found in \citet{collister}.

\subsection{Photometric redshift evaluation for the 2SLAQ sample}
\label{sec.photoz_eval}

Although the photometric target sample is expected to contain stellar contamination, we remove stars from the 2SLAQ sample (using their spectroscopic identification) before any photometric redshift estimation. As a result, ANNz does not recognise stars at all and assigns them extragalactic redshifts, but our tests show that the photometric redshift accuracy for the genuine galaxies is optimised by this approach. The stellar contamination is discussed in more detail in Section \ref{sec.sg}. Removing the stars reduces the 2SLAQ sample to 12,515 objects. 

In order to allow the photometric redshift accuracy to be assessed objectively, we separate out 8,515 2SLAQ objects at random to be used solely as an evaluation set. During the evaluation phase only the remaining 4,000 objects are used for training ANNz, and the evaluation set is treated strictly as a mock target sample. Figures \ref{fig3} and \ref{fig4} show that the evaluation set may be considered to be accurately representative of our photometric target sample (although caution is exercised at $i_\rmn{deV} > 19.8$ and $d_\rmn{perp} < 0.55$ where there are relatively few training objects; see Section \ref{sec.lowcomp}). We find that a 4,000-member training set is large enough to ensure convergence in terms of the photometric redshift accuracy: virtually identical results are obtained using a training set 8,000 members, but reducing the training set to 2,000 members results in a $\sim 5$ per cent increase in the overall photometric redshift error.

For photometric redshift estimation we use the model magnitudes in $griz$ as the inputs to ANNz. The model magnitudes are preferred due to their use of equal apertures in each band; this allows unbiased colour estimates, crucial to accurate photometric redshifts. We do not make use of the $u$ band primarily due to concern over a time-varying red leak in this filter (SDSS DR4 web site) that could introduce systematic coordinate dependence into the photometric redshift errors. Irrespective of this concern, the low signal to noise in the $u$ band relative to the other SDSS filters for LRGs means that it is found to contribute no measurable benefit to the photometric redshift accuracy. 

An ANNz committee of four networks with a 4:10:10:1 architecture (see \citealt{collister} for an explanation of this notation) was trained using the 4,000 object training set (split equally into training and validation subsets). Each member of the 8,515 object evaluation set was then submitted to the trained committee in order to obtain the photometric redshifts. 

\begin{figure}
\includegraphics[width=0.47\textwidth]{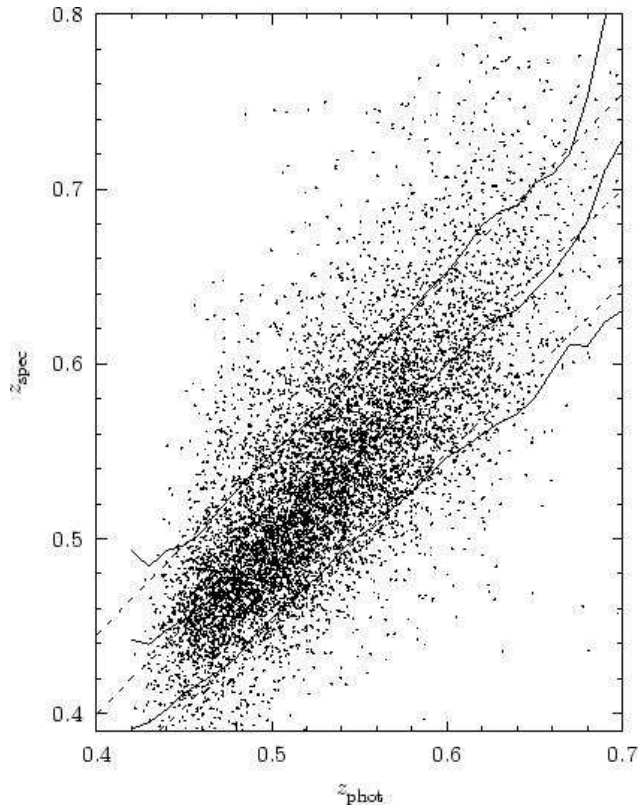}
\caption{Photometric redshift accuracy for the LRG evaluation set described in Section \ref{sec.photoz_eval}. The dashed lines show $z_\rmn{spec} = z_\rmn{phot}$ and the interval $\pm\sigma_0 (1 + z_\rmn{phot})$. The solid lines show the mean and standard deviation of $z_\rmn{spec}$ as a function of $z_\rmn{phot}$, evaluated in bins of $z_\rmn{phot}=0.02$.}
\label{fig5}
\end{figure}

The photometric redshifts for the evaluation set are plotted against the spectroscopic redshifts in Figure \ref{fig5}. The photometric redshift accuracy is characterised by the bias,$\langle \delta z\rangle$, where
\begin{equation}
\delta z \equiv z_\rmn{phot} - z_\rmn{spec}
\end{equation}
and the dispersion
\begin{equation}
\sigma^2_z = \left<(\delta z)^2\right> - \left<\delta z\right>^2,
\end{equation}
or with the inevitable loss of accuracy due to the stretching of the spectrum at increased redshift factored out,
\begin{equation}
\sigma^2_0=\left<\left(\frac{\delta z}{1+z_\rmn{spec}}\right)^2\right> - \left<\frac{\delta z}{1+z_\rmn{spec}}\right>^2.
\end{equation}
Note that these are purely statistical measures which can only be calculated for the evaluation set.

\begin{figure}
\includegraphics[width=0.47\textwidth]{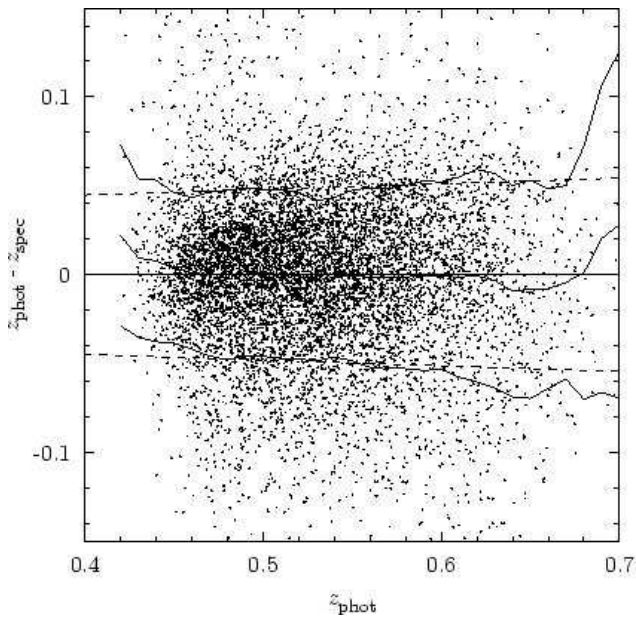}
\caption{Photometric redshift residuals versus photometric redshift. The dashed lines show the interval $\pm\sigma_0 (1 + z_\rmn{phot} )$ and the solid lines show the standard deviation of $z_\rmn{phot} - z_\rmn{spec}$.}
\label{fig6}
\end{figure}

For the evaluation set, the average photometric redshift error is $\sigma_z = 0.0488$ or $\sigma_0 = 0.0320$, similar to that obtained for the same sample by \citet{paddy}. Figure \ref{fig6} shows how the photometric redshift error varies with $z_\rmn{phot}$: the ANNz photometric redshift is seen to be an unbiased estimator for $z_\rmn{spec}$ over the range $0.45 < z_\rmn{phot} < 0.65$, with $|\langle\delta z\rangle| < 10^{-3}$, and the dispersion increases with redshift by no more than the expected factor of $(1+z)$. Very few objects are assigned $z_\rmn{phot} < 0.45$, and the average photometric redshift in this region is biased to higher values. At $z_\rmn{phot} > 0.65$ the dispersion noticeably worsens and the average photometric redshift is positively biased. 

It is possible, using ANNz, to estimate the contribution to the error budget originating from the photometric noise (as described in \citealt{collister}). Unsurprisingly, this contribution increases with redshift: at $z_\rmn{spec}<0.5$ the photometric noise is responsible for an average photometric redshift uncertainty of $\sigma_z = 0.021$, but this increases to $\sigma_z = 0.031$ for objects at $z_\rmn{spec}>0.6$. We also note that the average scatter between the outputs of the individual ANNs in the committee is $8 \times 10^{-4}$.

\begin{figure}
\includegraphics[width=0.47\textwidth]{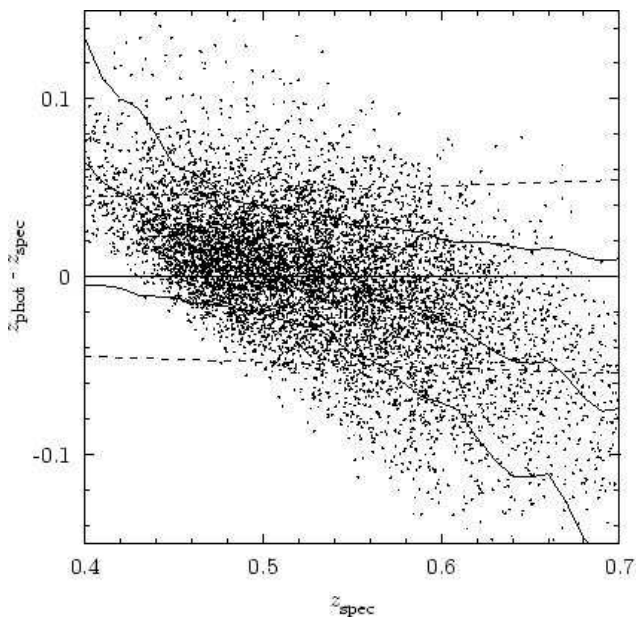}
\caption{As Figure \ref{fig6}, but showing the photometric redshift residuals versus spectroscopic redshift.}
\label{fig7}
\end{figure}

Figure \ref{fig7} shows the photometric redshift residuals as a function of the spectroscopic redshift, and makes it clear that the photometric redshift distribution is skewed with respect to the spectroscopic redshifts. Objects with lower spectroscopic redshifts tend to be assigned higher photometric redshifts on average, and objects with higher spectroscopic redshifts are assigned lower photometric redshifts. Note that this bias is not relevant when using the photometric redshifts to assign galaxies to bins: the mean spectroscopic redshift in a photometric redshift-selected bin will be close to the central photometric redshift in that bin, and the distribution around the mean is approximately symmetric. 

\begin{figure}
\includegraphics[width=0.47\textwidth]{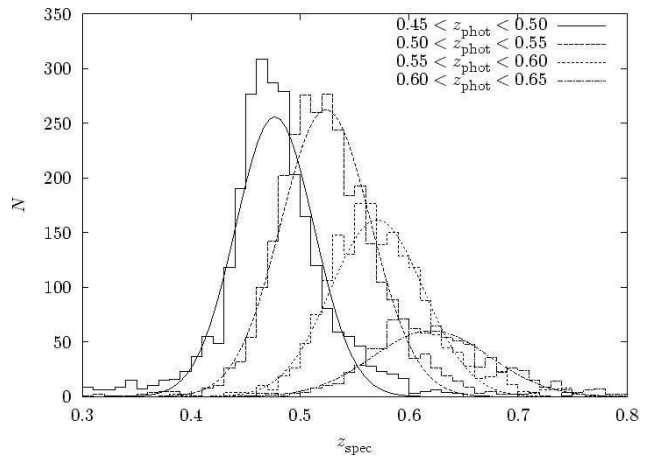}
\caption{Spectroscopic redshift distributions in photometric redshift-selected bins of width $\Delta z_\rmn{phot} = 0.05$.}
\label{fig8}
\end{figure}

Regardless of these general observations, it is critical in any application of the catalogue that a careful, individual assessment of the impact of the photometric redshift errors is made. As an example, we show the spectroscopic redshift distributions in photometric redshift-selected bins of width $\Delta z_\rmn{phot} = 0.05$ (Figure \ref{fig8}). Gaussian fits to these distributions are overplotted, and the mean and variance of the fits are given in Table \ref{tab.gauss}. The Gaussian provides a reasonable fit to the distribution in each of the bins. However, particularly in the $0.45 < z_\rmn{phot} < 0.50$ bin, the actual distribution has a tighter core and stronger wings than the fitted Gaussian. \citet{paddy} obtain similar distributions from their independent photometric redshift estimation. Analyses based on these bins should take care that this non-Gaussianity is taken into account: ideally one should use the measured $n(z_\rmn{spec})$ directly, rather than attempting to parameterize the distribution. The evaluation set photometric and spectroscopic redshifts have been made available with the MegaZ-LRG catalogue for this specific purpose.

\begin{table}
\caption{Mean and variance of Gaussian distributions fitted to the spectroscopic redshift distributions in photometric redshift selected bins (Figure \ref{fig8})}. 
\label{tab.gauss}
\centering
\begin{tabular}{crr}
\hline
& $\mu$ & $\sigma$\\
\hline 
$0.45 < z_\rmn{phot} < 0.50$ & 0.477 & 0.036\\
$0.50 < z_\rmn{phot} < 0.55$ & 0.524 & 0.041\\
$0.55 < z_\rmn{phot} < 0.60$ & 0.571 & 0.043\\
$0.60 < z_\rmn{phot} < 0.65$ & 0.620 & 0.054\\
\hline
\end{tabular}
\centering
\end{table}

It is important to note that these spectroscopic redshift distributions are not expected to be appropriate to samples selected from the complete MegaZ-LRG catalogue; they may only be used if cuts are applied at $i_\rmn{deV} = 19.8$ and $d_\rmn{perp} = 0.55$ to bring the MegaZ-LRG selection into line with the 2SLAQ sample (see Section \ref{sec.criteria}). 

\begin{figure}
\includegraphics[width=0.47\textwidth]{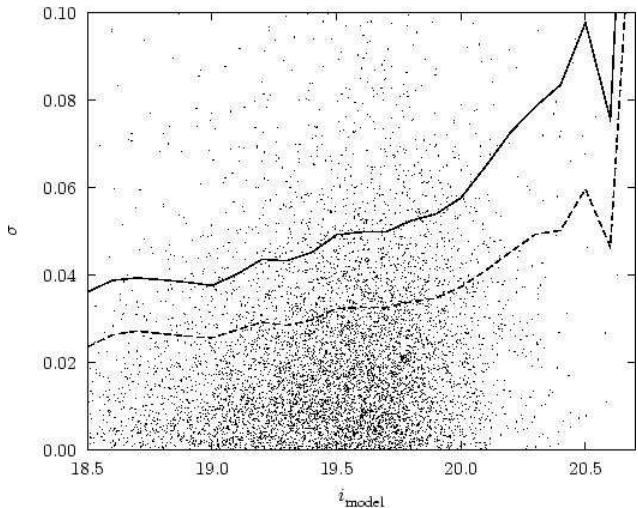}
\caption{Dependence of the photometric redshift accuracy on the $i$-band model magnitude. Points show the individual redshift errors for galaxies in the evaluation set. The solid line shows $\sigma_z$ and the dashed line shows $\sigma_0$ (which may be considered a fairer measure since it corrects for the different redshift distributions of samples defined by different magnitude ranges).}
\label{fig9}
\end{figure}

Figure \ref{fig9} displays the dependence of the photometric redshift accuracy on the $i$-band model magnitude. The redshift error increases steadily until $i_\rmn{model}\sim20$, beyond which the accuracy degrades much more rapidly as the limiting magnitude is approached. The photometric redshift accuracy for the brightest objects is significantly better than the average: considering only objects with $i_\rmn{model} < 19.0$, we find $\sigma_z = 0.0400$ and $\sigma_0 = 0.0276$.

\subsubsection{Accuracy in low-completeness training regions}
\label{sec.lowcomp}

As explained above, the 2SLAQ catalogue suffers from low completeness towards certain limits of the parameter space defined by the cuts in Section \ref{sec.selection}. The regions of interest can be seen in Figures \ref{fig3} and \ref{fig4} to host large numbers of target objects, but relatively few training objects. We now assess whether the scarcity of training data in these regions adversely impacts the photometric redshift accuracy.

\begin{figure}
\includegraphics[width=0.47\textwidth]{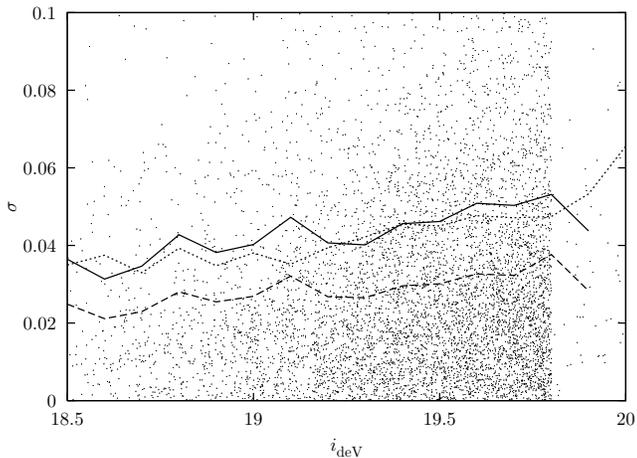}
\caption{As Figure \ref{fig9} but showing the dependence of the photometric redshift accuracy on the $i$-band de Vaucouleurs magnitude. The low sensitivity of $\sigma$ to $i_\rmn{deV}$ may be attributed to the significant stochasticity between $i_\rmn{deV}$ and $i_\rmn{model}$ (Figure \ref{fig11}). To allow more direct comparison, the dotted line shows $\sigma_z$ for $i_\rmn{model}$.}
\label{fig10}
\end{figure}

\begin{figure}
\includegraphics[width=0.47\textwidth]{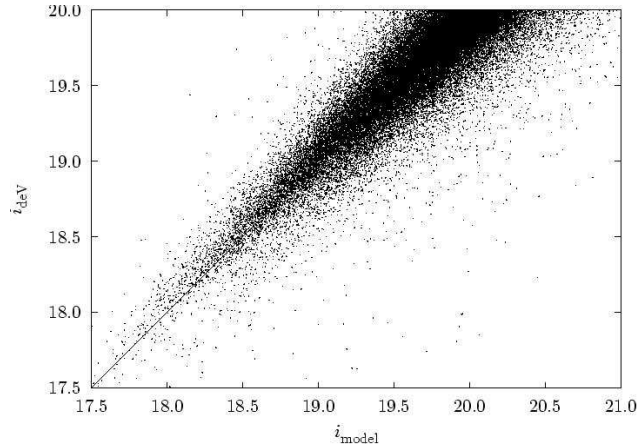}
\caption{Comparison of the $i$-band de Vaucouleurs and model magnitudes for the MegaZ-LRG sample.}
\label{fig11}
\end{figure}

We first examine the photometric redshift errors for objects having $19.8 < i_\rmn{deV} < 20.0$. In fact, as Figure \ref{fig10} shows, the photometric redshift accuracy in this regime shows no suggestion of deterioration. For the faintest objects, the photometric redshift accuracy is a much stronger function of the model magnitudes since it is these which are passed to ANNz and used to compute galaxy colours. The significant stochasticity between the $i$-band de Vaucouleurs and model magnitudes (Figure \ref{fig11}) ensures that this dependence is not propagated to $i_\rmn{deV}$, and the photometric redshift accuracy in $19.8 < i_\rmn{deV} < 20.0$ is close to the average for the sample as a whole. Note that $i_\rmn{model}$ is determined using the best-fitting profile in the $r$-band, whereas $i_\rmn{deV}$ uses the de Vaucouleurs model fitted in the $i$-band; these are not equal, in general, even when the model magnitude is based on a de Vaucouleurs profile . See \citet{york} for details. 

A similar concern applies to the $d_\rmn{perp} < 0.55$ region. The photometric redshift error evaluated only on these objects is $\sigma_z = 0.051$, $\sigma_0 = 0.033$; a little higher than the average but still an acceptable error level. Note that $d_\rmn{perp}$ is an excellent redshift indicator: lower $d_\rmn{perp}$ implies lower redshift. These $0.50 < d_\rmn{perp} < 0.55$ objects typically have $z_\rmn{spec}\sim0.45$.

\subsection{Catalogue photometric redshifts}

The photometric redshifts for the final MegaZ-LRG catalogue were estimated using a new ANNz committee of four 4:10:10:1 networks, trained on the entire 2SLAQ catalogue (except for the spectroscopically-identified stars): 12,515 objects, equally assigned to the training and validation subsets.

\begin{figure}
\includegraphics[width=0.47\textwidth]{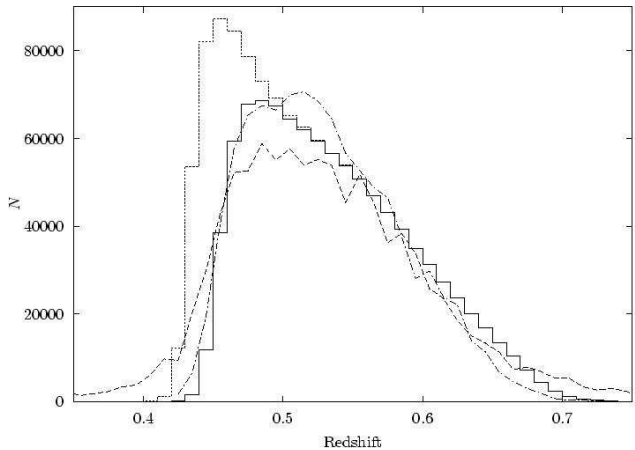}
\caption{Redshift distributions: (dotted histogram) photometric redshift distribution of the MegaZ-LRG catalogue; (solid histogram) photometric redshift distribution of the MegaZ-LRG catalogue including only objects with $d_\rmn{perp} > 0.55$; (dashed line) spectroscopic redshift distribution of the 2SLAQ evaluation set; (dot-dashed line) photometric redshift distribution of the 2SLAQ evaluation set. The evaluation set distributions are normalised to have the same integrated area as the MegaZ-LRG $d_\rmn{perp} > 0.55$ histogram. The histograms have bin width $\Delta z = 0.01$.}
\label{fig12}
\end{figure}

Figure \ref{fig12} shows the photometric redshift distribution of the MegaZ-LRG catalogue together with both the spectroscopic and photometric redshift distributions of the 2SLAQ evaluation sample. The MegaZ-LRG catalogue contains considerably more objects around $z_\rmn{phot}\sim0.45$ than the 2SLAQ evaluation sample. This is due to the admittance of objects having $0.50 < d_\rmn{perp} < 0.55$: there are 228,520 such objects in the catalogue, hence the significant boost to numbers at lower redshifts. If these objects are removed, the photometric redshift distributions of the MegaZ-LRG catalogue and 2SLAQ evaluation set are very similar.

\section{Enhanced star/galaxy separation}
\label{sec.sg}

Neural networks have previously been successfully applied to the star/galaxy separation problem by \citet{bert}, and the ANNz photometric redshift code can be effectively and straightforwardly applied to the task. We have trained ANNz to perform star/galaxy separation using the 2SLAQ catalogue as a training set. Instead of using the spectroscopic redshift as the target network output, we define $\delta_\rmn{sg}$ , such that $\delta_\rmn{sg}$ = 1 if the training object is a galaxy, and $\delta_\rmn{sg}$ = 0 if it has been spectroscopically identified as a star. The output of an ANN trained to predict $\delta_\rmn{sg}$ will be a continuous quantity; it can be shown that this output may be interpreted as the classification probability for the particular target object. The closer $\delta_\rmn{sg}$ is to 1, the higher the probability that the particular target object is a galaxy. 

The ANN method has two especially attractive advantages: (i) we can allow the network to consider as many parameters as we believe may be relevant to the problem of star/galaxy separation, and (ii) we do not need to construct \emph{ad hoc} criteria such as those of Section \ref{sec.selection}, but can simply leave it to the network to determine the optimal classification scheme. 

We selected 15 SDSS photometric parameters to be used as inputs to the star/galaxy classifying ANN; these are listed in Table \ref{tab.sgpars}. They include the object's magnitude in each of the SDSS $griz$ filters (the $u$ band is not used for the reasons outlined in Section \ref{sec.photoz}), along with a number of parameters describing the angular size and the distribution of light within the object. Other parameters were considered, in particular the Petrosian magnitudes and radii, but these were found to result in negligible improvement in the separation. As in Section \ref{sec.photoz}, we separated out an evaluation set of 9,139 objects, now selected from 13,139 2SLAQ objects since the stars are included. A committee of four 15:20:20:1 networks was trained to predict $\delta_\rmn{sg}$ using the remaining 4,000 objects as the training set. The trained committee was then applied to the evaluation set, in order to obtain the predicted galaxy probability for each object. The initial stellar contamination fraction in both the training and evaluation samples was 5 per cent.

\begin{table}
\caption{Photometric parameters used as inputs to ANNz for star/galaxy separation. Apart from the dereddened model magnitudes and the SDSS type classification, all parameters are measured in the i band, since these red objects exhibit greatest signal-to-noise in this filter. Note that only the four dereddened model magnitudes were used during the photometric redshift estimation.}
\label{tab.sgpars}
\centering
\begin{tabular}{ll}
\hline
Parameter & Description \\
\hline
\texttt{dered\_g} &\\
\texttt{dered\_r} & Dereddened model magnitudes\\
\texttt{dered\_i} &\\
\texttt{dered\_z} &\\
\texttt{psfMag\_i} & PSF flux (dereddened)\\
\texttt{fiberMag\_i} & Flux in 3\arcsec diameter fibre radius (dereddened)\\
\texttt{deVMag\_i} & De Vaucouleurs magnitude (dereddened)\\
\texttt{expMag\_i} & Exponential fit magnitude (dereddened)\\
\texttt{deVRad\_i} &De Vaucouleurs fit scale radius\\
\texttt{deVAB\_i} &De Vaucouleurs fit axis ratio\\
\texttt{expRad\_i} & Exponential fit scale radius\\
\texttt{expAB\_i} &Exponential fit axis ratio\\
\texttt{lnLStar\_i} &Star ln(likelihood)\\
\texttt{lnLExp\_i} &Exponential disk fit ln(likelihood)\\
\texttt{lnLDeV\_i} &DeVaucouleurs fit ln(likelihood)\\
\hline
\end{tabular}
\centering
\end{table}

\begin{figure}
\includegraphics[width=0.47\textwidth]{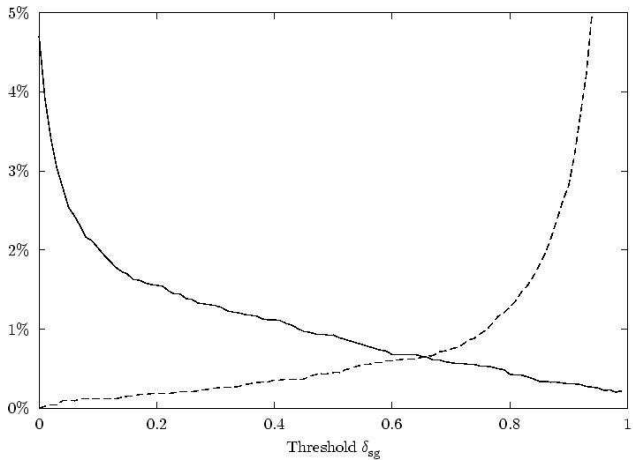}
\caption{Effect of varying the ANNz star/galaxy separation threshold. The solid line shows the number of stars passing the cut as a fraction of the total number of objects admitted. The dashed line shows the fraction of genuine galaxies which are discarded, as a fraction of the total number of genuine galaxies in the original sample.}
\label{fig13}
\end{figure}

In order to perform separation one must decide on a threshold probability for admittance to the galaxy sample. Increasing this threshold leads to more aggressive removal of stars, but may also cause more genuine LRGs to be discarded. Figure \ref{fig13} shows the effect of varying the admittance threshold on the contamination level in the evaluation set. A conservative suggestion would be to adopt a threshold galaxy probability of 0.2: this is expected to reduce the expected stellar contamination to 2 per cent, with the loss of only 0.1 per cent of the genuine galaxies. Alternatively, a more agressive threshold of 0.8 reduces the expected stellar contamination to just 0.5 per cent, but still preserves all but $\sim$1 per cent of the genuine galaxies.

To obtain $\delta_\rmn{sg}$ values for each of the objects in the MegaZ-LRG catalogue, a new committee of four 15:20:20:1 networks was trained, now using the entire 2SLAQ catalogue as the training set (13,139 objects split equally into training and validation subsets). Applying an admittance threshold of 0.2 to the MegaZ-LRG catalogue reduces its size to 1,190,682 objects. Note that the MegaZ-LRG catalogue has not had any such cut applied, but $\delta_\rmn{sg}$ is provided for each object. Star/galaxy separation may be performed by discarding objects having $\delta_\rmn{sg}$ less than one's preferred threshold (guided by Figure \ref{fig13}).

\section{Summary: the MegaZ-LRG catalogue}
\label{sec.megazlrg}

We have selected a photometric sample from the SDSS DR4 imaging catalogue using the criteria devised for the 2SLAQ LRG survey. This MegaZ-LRG catalogue contains 1,214,117 objects in total. Luminous red galaxies are expected to comprise $\sim95$ per cent of the catalogue membership, with the remaining $\sim$5 per cent dominated by M-type stars. 

The 2SLAQ spectroscopic catalogue of  $\sim$13,000 LRGs was used to train the ANNz photometric redshift code. For each of the catalogue objects, photometric redshifts were estimated based on the dereddened $griz$ model magnitudes. The rms photometric redshift error computed for an evaluation set selected from the 2SLAQ sample is $\sigma_z = 0.049$, or $\sigma_0 = 0.030$. 

ANNz was separately trained to perform star/galaxy separation based on a set of 15 photometric parameters. A star/galaxy flag, $\delta_\rmn{sg}$, was estimated for each of the catalogue objects; this continuous parameter may be interpreted as the probability that a particular object is a galaxy rather than a star. The aggressiveness of star/galaxy separation may be varied through the choice of threshold imposed on the $\delta_\rmn{sg}$ parameter. 

The MegaZ-LRG catalogue may be obtained from \texttt{http://www.2slaq.info}. For each of the 1,214,117 objects in the catalogue we provide the photometric redshift and $\delta_\rmn{sg}$ parameter calculated as described above. To allow any of the full range of SDSS photometric parameters to be straightforwardly obtained we also include each object's SDSS \texttt{objID}. A basic set of photometric parameters are included with the catalogue for convenience (Table \ref{tab.catpars}). An example listing of six objects from the MegaZ-LRG catalogue is provided in Table \ref{tab.sample}.

\begin{table}
\caption{Parameters included in the MegaZ-LRG photometric redshift catalogue.}
\label{tab.catpars}
\centering
\begin{tabular}{ll}
\hline 
\texttt{objID} & SDSS objID\\
\texttt{ra} &J2000 right ascension \\
\texttt{dec} &J2000 declination\\
\texttt{dered\_u}&\\ 
\texttt{dered\_g}&\\
\texttt{dered\_r} &Dereddened model magnitudes\\
\texttt{dered\_i} &\\
\texttt{dered\_z} &\\
\texttt{deVMag\_i} &Dereddened de Vaucouleurs magnitude\\
\texttt{z\_phot} & ANNz photometric redshift\\
\texttt{delta\_sg}& ANNz galaxy probability\\
\hline
\end{tabular}
\centering
\end{table}

\begin{table*}
\caption{Example extract of six objects from the MegaZ-LRG catalogue.}
\label{tab.sample}
\centering
\begin{tabular}{lrrrrrrrrrr}
\hline
\texttt{objID} &\texttt{ra} &\texttt{dec} &\texttt{dered\_u} &\texttt{dered\_g}& \texttt{dered\_r} &\texttt{dered\_i}& \texttt{dered\_z}& \texttt{deVMag\_i}& \texttt{z\_phot}& \texttt{delta\_sg}\\
\hline
587722952230174879 &236.309 &-0.430 &22.482 &21.051 &19.339 &18.540 &18.119 &18.392 &0.455 &0.987\\
587722952230175340 &236.338 &-0.579 &22.864 &22.350 &21.209 &19.972 &19.584 &19.781 &0.644 &0.987\\
587722952230175431 &236.246 &-0.449 &22.171 &22.449 &21.089 &19.965 &19.643 &19.613 &0.600 &0.998\\
587722952230175557 &236.299 &-0.573 &24.406 &21.805 &20.501 &19.714 &19.184 &19.955 &0.514 &0.266\\
587722952230175583 &236.307 &-0.436 &24.250 &22.456 &20.886 &19.793 &19.613 &19.956 &0.570 &0.999\\
587722952230175590 &236.311 &-0.513 &22.855 &22.185 &20.872 &20.099 &19.720 &19.973 &0.513 &0.999\\
\hline
\end{tabular}
\centering
\end{table*}

Measurements of large-scale structure within photometric redshift slices in the MegaZ-LRG catalogue are presented in \citet{blake}. An independent analysis by \citet{paddy06}, based on a similar sample, has produced consistent results . A study of non-linear clustering within MegaZ-LRG \citep{phd} is to appear shortly, and the MegaZ-LRG catalogue is due to be extended to make use of the additional area provided in the SDSS Data Release 5.

\section*{Acknowledgements}

We thank Filipe Abdalla and Sarah Bridle for helpful comments and discussion during this work, and all members of the AAO staff who helped to run and maintain 2dF during the course of the 2SLAQ survey.

AC was supported by an Isle of Man Department of Education Postgraduate Studies Grant. OL acknowledges a PPARC Senior Research Fellowship. CB acknowledges support from the Izaak Walton Killam Memorial Fund for Advanced Studies and from the Canadian Institute for Theoretical Astrophysics National Fellowship programme. 

Funding for the SDSS and SDSS-II has been provided by the Alfred P. Sloan Foundation, the Participating Institutions, the National Science Foundation, the U.S. Department of Energy, the National Aeronautics and Space Administration, the Japanese Monbukagakusho, the Max Planck Society, and the Higher Education Funding Council for England. The SDSS Web Site is \texttt{http://www.sdss.org/}. 

The SDSS is managed by the Astrophysical Research Consortium for the Participating Institutions. The Participating Institutions are the American Museum of Natural History, Astrophysical Institute Potsdam, University of Basel, Cambridge University, Case Western Reserve University, University of Chicago, Drexel University, Fermilab, the Institute for Advanced Study, the Japan Participation Group, Johns Hopkins University, the Joint Institute for Nuclear Astrophysics, the Kavli Institute for Particle Astrophysics and Cosmology, the Korean Scientist Group, the Chinese Academy of Sciences (LAMOST), Los Alamos National Laboratory, the Max-Planck-Institute for Astronomy (MPIA), the Max-Planck-Institute for Astrophysics (MPA), New Mexico State University, Ohio State University, University of Pittsburgh, University of Portsmouth, Princeton University, the United States Naval Observatory, and the University of Washington.

\label{lastpage}

\begin{thebibliography}{}
\bibitem[Adelman-McCarthy et al.(2006)]{adel}
  Adelman-McCarthy, J. K., et al. 2006, ApJS, 162, 38
\bibitem[Ben\'itez(2000)]{benitez}
  Ben\'itez N., 2000, ApJ, 536, 571
\bibitem[Bertin \& Arnouts(1996)]{bert}
  Bertin E., Arnouts S., 1996, A\&AS, 117, 393 
\bibitem[Blake et al.(2006)]{blake}
  Blake C. A., Collister A. A., Bridle S. L., Lahav O., 2006, preprint (astro-ph/0605303)
\bibitem[\protect\citeauthoryear{Bolzonella, Miralles \& Pell\'o}{Bolzonella et~al.}{2000}]{bolz}
  Bolzonella M., Miralles J.-M., Pell\'o R., 2000, A\&A, 363, 476 
\bibitem[Cannon et al.(2006)]{cannon}
  Cannon R. D., et al., 2006, submitted to MNRAS 
\bibitem[Colless et al.(2001)]{colless}
  Colless M., et al., 2001, MNRAS, 328, 1039 
\bibitem[Collister(2006)]{phd}
  Collister A. A., 2006, PhD thesis, Cambridge University 
\bibitem[Collister \& Lahav(2004)]{collister}
  Collister A. A., Lahav O., 2004, PASP, 116, 345
\bibitem[Connolly et al.(1995)]{connolly}
  Connolly A. J., Csabai I., Szalay A. S., Koo D. C., Kron R. G., Munn J. A., 1995, AJ, 110, 2655
\bibitem[Csabai et al.(2003)]{csabai}
  Csabai I. et al., 2003, AJ, 125, 580 
\bibitem[Eisenstein et al.(2001)]{eis01}
  Eisenstein D. J., Annis J., Gunn J. E., et al., 2001, AJ, 122, 2267 
\bibitem[Eisenstein et al.(2003)]{eis03}
  Eisenstein D. J., et al., 2003, ApJ, 585, 694 
\bibitem[\protect\citeauthoryear{Firth, Lahav \& Somerville}{Firth et~al.}{2003}]{firth}
  Firth A. E., Lahav O., Somerville R. S., 2003, MNRAS, 339, 1195 
\bibitem[Fukugita et al.(1996)]{fuku}
  Fukugita M., Ichikawa T., Gunn J. E., Doi M., Shimasaku K., Schneider D. P., 1996, AJ, 111, 1748
\bibitem[Gunn et al.(1998)]{gunn98}
  Gunn J. E., Carr M. A., Rockosi C. M., Sekiguchi M., et al., 1998, AJ, 116, 3040 
\bibitem[Gunn et al.(2006)]{gunn06}
  Gunn J. E., Siegmund W. A., Mannery E. J., Owen R. E., et al., 2006, AJ, 131, 2332 
\bibitem[Padmanabhan et al.(2005)]{paddy}  
  Padmanabhan N., Budav\'ari T., Schlegel D., et al., 2005, MNRAS, 359, 237
\bibitem[Padmanabhan et al.(2006)]{paddy06}
  Padmanabhan N., et al., 2006, preprint (astro-ph/0605302)
\bibitem[Pier et al.(2003)]{pier}
  Pier J. R., Munn J. A., Hindsley R. B., Hennessy G. S., Kent S. M., Lupton R. H., Ivezic Z., 2003, AJ, 125, 1559
\bibitem[\protect\citeauthoryear{Schlegel, Finkbeiner \& Davis}{Schlegel et~al.}{1998}]{schlegel}
Schlegel D. J., Finkbeiner D. P., Davis M., 1998, ApJ, 500, 525
\bibitem[Smith et al.(2002)]{smith}
  Smith J. A., Tucker D. L., Kent S. M., et al., 2002, AJ, 123, 2121
\bibitem[Stoughton et al.(2002)]{stoughton}
  Stoughton C., Lupton R. H., Bernardi M., et al., 2002, AJ, 123, 485
\bibitem[York et al.(2000)]{york}
  York, D.G., et al. 2000, AJ, 120, 1579
\end{thebibliography}
\end{document}